\documentclass[twocolumn]{aastex631}

\begin{document}

\title{X-ray and radio polarimetry of the neutron star low-mass X-ray binary 4U~1728$-$34}

\author{Unnati Kashyap}
\affiliation{Department of Physics and Astronomy, Texas Tech University, Lubbock, TX 79409-1051, USA} 

\author{Thomas J. Maccarone}
\affiliation{Department of Physics and Astronomy, Texas Tech University, Lubbock, TX 79409-1051, USA}

\author{Thomas D. Russell}
\affiliation{INAF, Istituto di Astrofisica Spaziale e Fisica Cosmica, Via U. La Malfa 153, I-90146 Palermo, Italy}

\author{Mason Ng}
\affiliation{Department of Physics, McGill University, 3600 rue University, Montr\'{e}al, QC H3A 2T8, Canada}
\affiliation{Trottier Space Institute, McGill University, 3550 rue University, Montr\'{e}al, QC H3A 2A7, Canada}

\author{Swati Ravi}
\affiliation{MIT Kavli Institute for Astrophysics and Space Research, Massachusetts Institute of Technology, Cambridge, MA 02139, USA}

\author{Eliot C. Pattie}
\affiliation{Department of Physics and Astronomy, Texas Tech University, Lubbock, TX 79409-1051, USA}

\author{Herman L. Marshall}
\affiliation{MIT Kavli Institute for Astrophysics and Space Research, Massachusetts Institute of Technology, Cambridge, MA 02139, USA}

\begin{abstract}
We report the first X-ray and radio polarimetric results of the neutron star (NS) low-mass X-ray binary (LMXB) atoll-source 4U~1728$-$34 using the Imaging X-ray Polarimetry Explorer (IXPE) and Australia Telescope Compact Array (ATCA). We discovered that the X-ray source was polarized at  ${\rm PD} = 1.9 \pm 1.0\%$  with a polarization angle of ${\rm PA} = -41 \pm16 \degr$. Simultaneous Neutron Star Interior Composition Explorer (NICER) observations show that the source was in a relatively hard state, marking it as the first IXPE observation of an NS atoll source in the hard state. We do not detect any significant linear polarization (LP) in the radio band, with a 3$\sigma$ upper limit of 2\% at 5.5\, GHz and 1.8\% at 9\, GHz. Combining the radio datasets provides the deepest upper limits on the radio polarization at $<$ 1.5\% on the linear and circular polarization (measured at 7.25\, GHz). The X-ray polarimetric results suggest a source geometry with a Comptonization component possibly attributed to a boundary layer (BL) emission reflected off the disk, consistent with the other NS atoll sources.

\end{abstract}


\section{Introduction} \label{sec:intro}
A low-mass X-ray binary (LMXB) is composed of a neutron star (NS) or a stellar-mass black hole (BH) that accretes matter from a low-mass companion star via Roche-lobe overflow \citep{2023hxga.book..120B}. A significant population of NS LMXBs is persistently accreting, exhibiting distinct states with varying accretion rates, with strongly correlated evolution in the source spectral and timing properties \citep{2001AdSpR..28..307B,2004astro.ph.10551V,2007A&ARv..15....1D,2007ApJ...667.1073L}. Based on the joint spectral and temporal nature, and the shape of the path that they trace out in the X-ray color-color diagram (CCD) or hardness intensity diagram (HID), they can be classified as: High Soft State (HSS) Z-sources ($>10^{38}$ erg/s), analogous to the Very High State/transition states in the BH X-ray binaries; HSS bright atoll sources ($10^{37}$ -- $10^{38}$ erg/s); and Low Hard State (LHS) atoll sources ($\sim 10^{36}$ erg/s) \citep{1989A&A...225...79H}. The Z-sources trace out Z-shaped tracks in the HIDs consisting of three branches, called the horizontal branch (HB), the normal branch (NB), and the flaring branch (FB). In contrast, the atoll sources trace a well-defined banana (bright atolls) and island state (LHS atolls) in the HIDs \citep{1995foap.conf..213V}. An evolution in radio jet emission (seen in the radio to mm band) is also observed during the transitions through different X-ray spectral states \citep{2006csxs.book..381F}.

 Typically, the X-ray spectra of NS LMXBs are dominated by thermal components in the soft states. In the hard state (sometimes referred to as the ``island state"), however, the spectra are often described by Compton-scattered blackbody radiation dominating in the hard X-rays, along with a directly observed blackbody/multicolor disk blackbody representing the softer X-ray emission component \citep{2006A&A...459..187P,2011A&A...529A.155C}. Based on the choice of the thermal and non-thermal components, there are two classical models that describe the geometry of NS LMXBs, often referred to as the Eastern model (after \citealt{1989PASJ...41...97M}) and the Western model (after \citealt{1988ApJ...324..363W}). In the Eastern model, the soft component is a multi-temperature black body produced by the accretion disk, and the hard component represents the emission from Comptonization occurring in a boundary layer (BL) between the disk and the NS  \citep{1984PASJ...36..741M,1989PASJ...41...97M,2001ApJ...547..355P} or a more vertically extended spreading layer (SL) around the NS \citep{2006MNRAS.369.2036S}. Alternatively, the Western model assumes that the soft component is a single-temperature black body arising from the NS surface, while the hard component originates from the Comptonization of disk photons by the hot energetic electrons of the corona \citep{1988ApJ...324..363W}. Both models deal with two different black body (or multicolor) distributed photon populations and a single Comptonizing region. More complex scenarios, such as a three-component model, including both NS and disk thermal emission plus Comptonization \citep{2007ApJ...667.1073L} and both (disk and NS) seed photon populations scattered in the corona, have also been proposed \citep{2011A&A...529A.155C}.

Which model accurately explains the accretion geometry of NS LMXBs is still a matter of debate, as the X-ray spectra of NS LMXBs are degenerate spectroscopically \citep{2001AdSpR..28..307B}.  The eastern-model framework, however, has gained stronger support from the Fourier-frequency resolved spectroscopy of atoll/Z sources. The SL spectrum resembles the Fourier-frequency resolved spectrum at frequencies of Quasi-Periodic Oscillations (QPOs) \citep{2003A&A...410..217G,2013MNRAS.434.2355R}. The presence of rapid variability associated with the BL rather than the disk, provides evidence for the ‘Eastern model’ with a lower temperature disk and hotter, Comptonized BL \citep{2006A&A...453..253R}. Furthermore, X-ray polarization measurements can also help disentangling the accretion geometry of LMXBs. The Imaging X-ray Polarimetry Explorer (IXPE) has so far targeted several NS LMXBs like Cyg X-2 \citep{2023MNRAS.519.3681F}, GX~9$+$9 \citep{2023AA...676A..20U}, XTE~J1701$-$462 \citep{2023MNRAS.525.4657J,2023A&A...674L..10C,2024MNRAS.tmp.1829G,2024arXiv240102658Y}, 4U~1820$-$30 \citep{2023ApJ...953L..22D}, 4U~1624$-$49 \citep{2024ApJ...963..133S},  GX~13$+$1 \citep{2024A&A...688A.170B}, Sco~X-1 \citep{2024ApJ...960L..11L}, GS~1826$-$238 \citep{2023ApJ...943..129C}, GX~3$+$1 \citep{2024arXiv241110353G}, GX~5$-$1 \citep{2024AA...684A.137F}, Ser~X-1 \citep{2024AA...690A.200U}, Cir~X-1 \citep{2024ApJ...961L...8R}, and GX~340$+$0 \citep{2024arXiv240519324B,2024arXiv241100350B}. Recent polarization results rule out the accretion disk itself as the primary source of polarized X-rays, and also rule out a geometry in which the BL is coplanar with the disk \citep{2023MNRAS.519.3681F}. Moreover, the X-ray polarization angle (PA) of Cyg X-2 appears to be aligned with the radio jet \citep{2023MNRAS.519.3681F}, suggesting SL at the NS surface  as the main source of the polarized radiation. However, recent studies of Sco X-1 show an X-ray PA misalignment with respect to the direction of the radio jet \citep{2024ApJ...960L..11L}. Thus, it is crucial to determine the position angle of radio jets with respect to the X-ray PA to disentangle the changing geometry along the different HID tracks in NS LMXBs, and to do this for a substantial sample of sources at different luminosities and hardnesses, as the geometry of the emission region may vary with time for any given source.

 The NS LMXB 4U~1728$-$34 is classified as an atoll-type X-ray binary based on its spectral and timing properties \citep{1989A&A...225...79H} and has long been suggested to be an ultra-compact X-ray binary based on an 11-min periodicity that is likely due to binary orbital modulation and properties of its Type I bursts that indicate likely helium composition for the accreted material \citep{2010ApJ...724..417G,2003ApJ...593L..35S}. It consists of a weakly magnetized NS accreting from a hydrogen-poor donor star \citep{2003ApJ...593L..35S, 2010ApJ...724..417G}. However, the orbital period estimations ($\sim$ 1.1h or
even $\sim$ 3h) based on the infrared observations of the delay between X-ray bursts and its reflection off the disk and companion contradicts the proposed ultra-compact nature and suggests a helium star companion \citep{2023MNRAS.525.2509V}. \cite{2000ApJ...542.1034D,2006A&A...458...21F} show that the broadband energy spectrum of 4U~1728$-$34 is well described by a Comptonization model along with the soft thermal blackbody component arising from the accretion disk. Strong aperiodic variabilities such as kHz QPOs \citep{1997ApJ...487L..77S}, often in pairs (lower and upper kHz QPOs) have also been reported from this source \citep{2001ApJ...561.1016M,2003MNRAS.345L..35M,2024arXiv241206644S}. 4U~1728$-$34 exhibits a broad iron emission line \citep{2000ApJ...542.1034D} along with a strong reflection component \citep{2019MNRAS.484.3004W}. It is one of the most extensively studied NS LMXBs showing type I X-ray bursts with its estimated distance ranging from 4.4 to 5.1 kpc \citep{2000ApJ...542.1034D,2001ApJ...551..907V,2003ApJ...590..999G,2019MNRAS.487.1626Q}, along with evidence of the presence of burst-disk interaction \citep{2017A&A...599A..89K}. The previous Very Large Array (VLA) observation of 4U~1728$-$34 at 4.86 GHz shows that the source exhibits a variable flux density ranging between 0.3 and 0.6 mJy \citep{1998A&A...332L..45M}. Moreover, \cite{2003MNRAS.342L..67M} report a positive correlation between the radio flux density at 8.46 GHz and the 2-10 keV X-ray flux representing the first evidence for a coupling between disk and jet in an atoll-type X-ray binary. A hard-to-soft state transition of 4U~1728$-$34, consistent with a jet emission with a break from optically thick to optically thin emission has also been reported \citep{2017A&A...600A...8D}.

In this paper, we analyze IXPE observations of the NS atoll LMXB 4U~1728$-$34 performed from 2024 Aug 17 to 2024 Aug 19. The simultaneous NICER observations with enhanced spectral sensitivity helped to constrain the spectral-timing-polarimetric properties of the
source. The Australia Telescope Compact Array (ATCA) observed 4U~1728$-$34 over three consecutive days in 2021 April. We extend the already existing analysis of those data \citep{2024Natur.627..763R} and obtain an upper limit of polarization of the radio jet of 4U~1728$-$34.

\section{X-ray observations}
\label{sec2}

\begin{figure}
\centering
\includegraphics[width=0.5\textwidth]{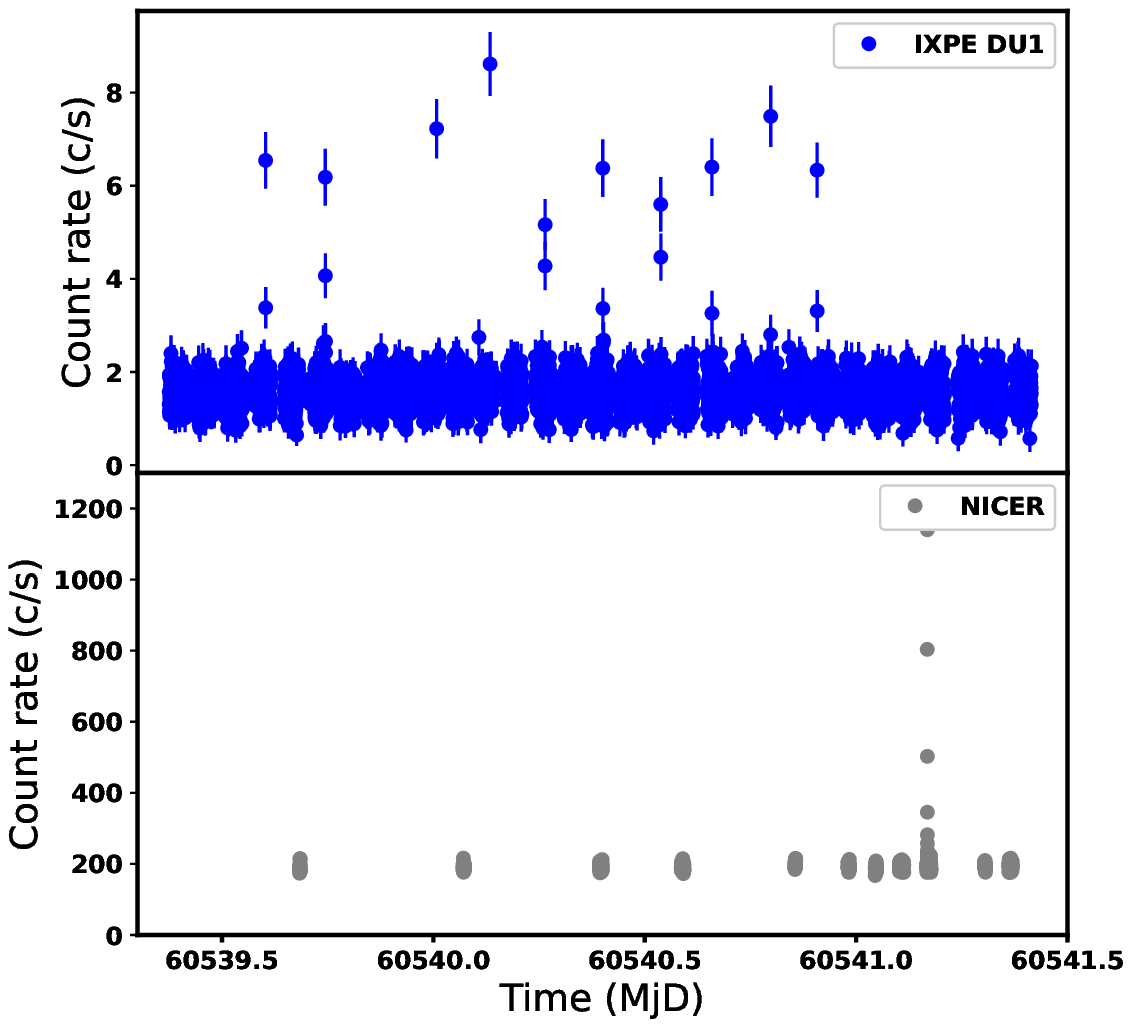}
\caption{Upper panel: IXPE (2-8 keV) background subtracted light curve of 4U~1728$-$34 (top). Time bins of 20 s are used. Lower Panel: NICER (0.5-10 keV) light curve of 4U~1728$-$34 (bottom). Time bins of 8 s are used.  }
\label{lc}
\end{figure}

\begin{figure}
\centering
\includegraphics[width=0.5\textwidth]{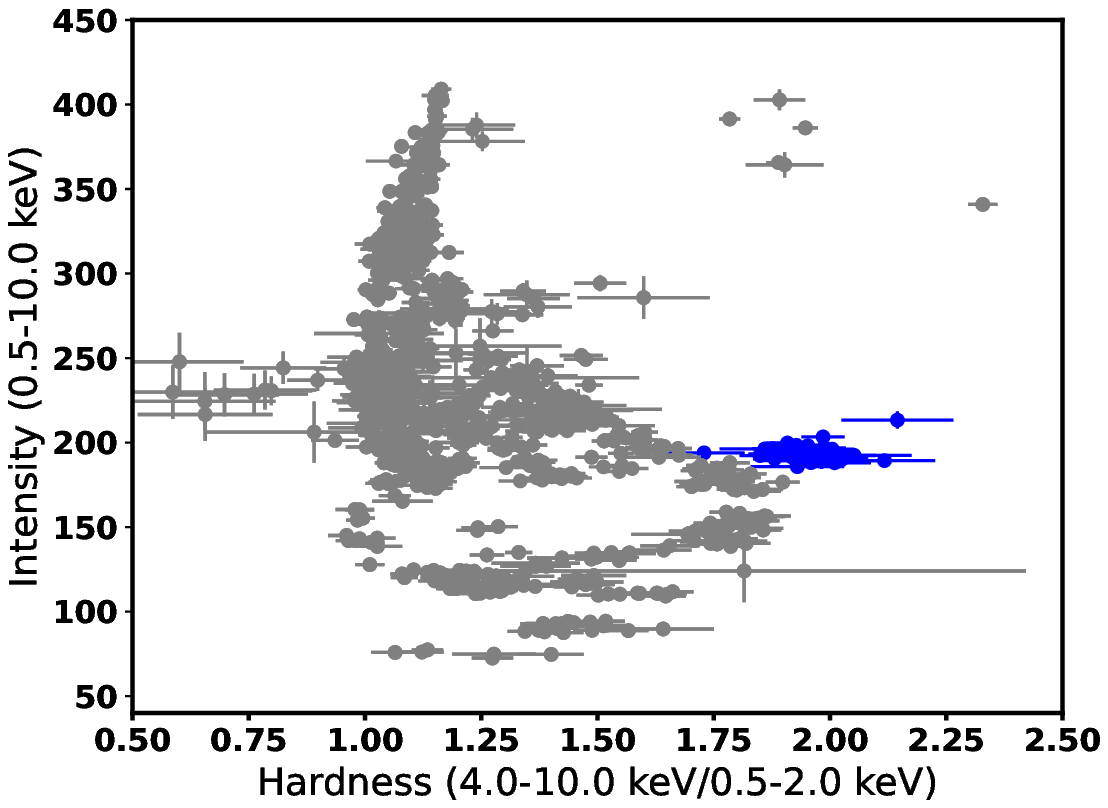}
\caption{Hardness–intensity diagram showing all NICER observations of
4U~1728$-$34 from 2017 June 29 to Sept 1, 2019. The hardness-intensity diagram from the current NICER observations of 4U~1728$-$34 is indicated by blue circles. Time bins of 128 s are used.}
\label{hid}
\end{figure}

\begin{figure}
\centering
\includegraphics[width=0.4\textwidth,angle=270]{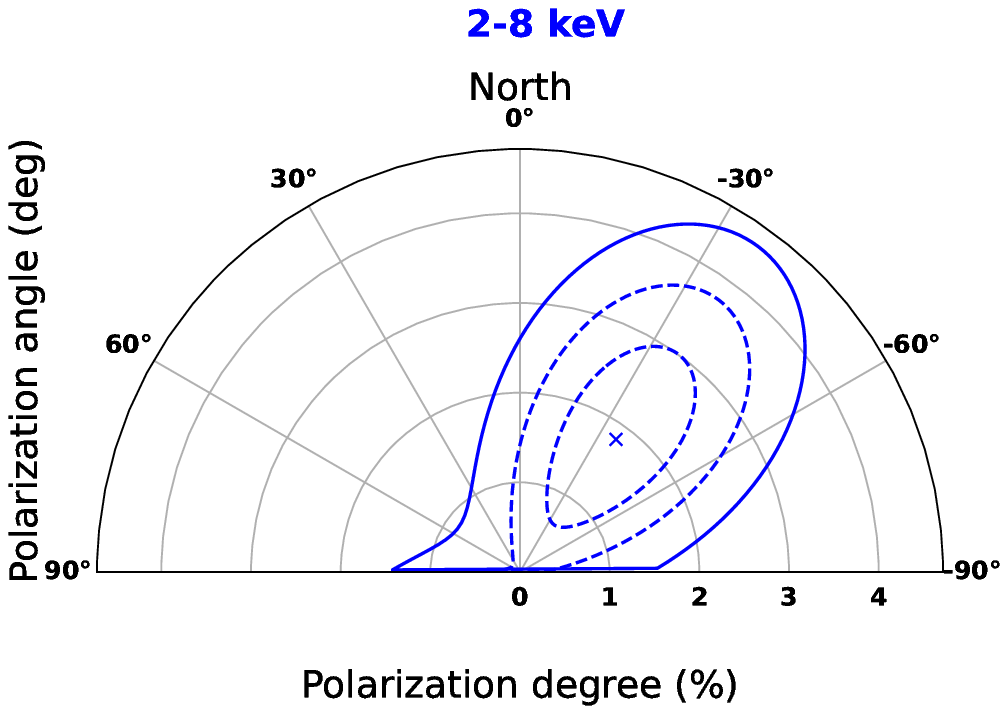}
\includegraphics[width=0.4\textwidth,angle=270]{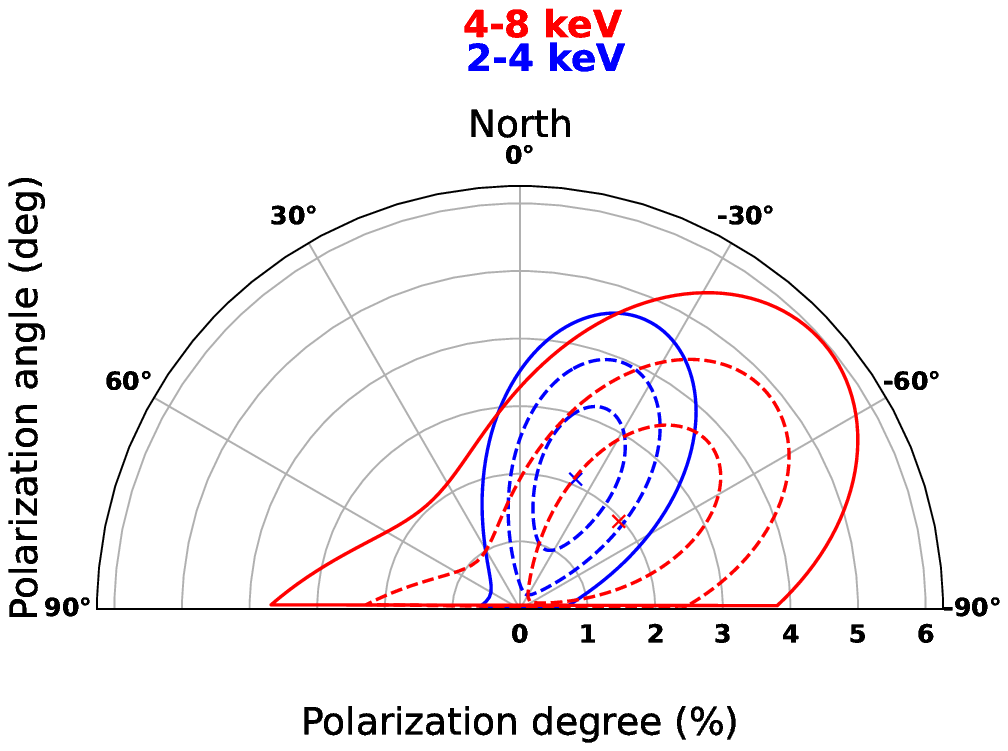}

\caption{Contour plots of the polarization degree and angle, determined with the {\tt PCUBE} algorithm, at the 68 \%, 95 \% and 99.7 \% confidence levels, in the 2–8 (upper panel, blue), 2-4 (lower panel, red) and 4–8 keV (lower panel, blue) energy bands.}
\label{polar_mod_ind}
\end{figure}

\begin{table*}
\centering
\caption{ {\em IXPE}, {\em NICER}, and {\em ATCA} Observations of 4U~1728$-$34 (see Section \ref{sec2}). }

\begin{tabular}{c c c c c}
\hline
Instrument & Observation ID & Date (dd-mm-yyyy) & Start time in UTC (hh:mm:ss) &Exposure time (ks)  \\ \hline

IXPE &03003701&17-08-2024--19-08-2024& 08:59:42  & 95.8 \\
NICER &7050150102-104&17-08-2024--19-08-2024 &16:21:26& 5.1\\
ATCA &&03-04-2021--04-04-2021&12:44:49&44.4  \\
&&04-04-2021--05-04-2021&13:17:40&42.4 \\
&&05-04-2021--05-04-2021&13:08:59& 20.9\\
\hline
\label{table1}
\end{tabular}

\end{table*}

\begin{table}
\label{table 2}
\centering
\caption{Results obtained from the {\tt PCUBE} analysis. The uncertainties are 1 $\sigma$. These estimates are for the background-rejected (and not background-subtracted) data (see Section \ref{model_ind_analysis}). }

\begin{tabular}{c c c c}
\hline
Energy band& PD (\%) & PA (\degr) \\ 
\hline

2-8 keV &1.8 $\pm$ 0.8 &-35 $\pm$ 12 \\
2-4 keV &2.1 $\pm$ 0.7&-23 $\pm$ 10 \\
4-8 keV &2.0 $\pm$ 1.2&-48 $\pm$ 17 \\

\hline
\end{tabular}

\end{table}

\subsection{IXPE}
\begin{table}
\centering

\caption{Best-fitting spectral model parameters from an absorbed disk-blackbody and Comptonization model {\tt tbnew*(diskbb+ compTT+ gaussian)*polconst*const} to the joint NICER and IXPE spectra of 4U~1728$-$34. The uncertainties are 1 $\sigma$ (see Section \ref{spec_pol_analysis}). The calibration constant for NICER is fixed at unity.}   

\begin{tabular}{c c c}
\hline
Parameter (unit)& &  \\ 
\hline
&tbnew&\\
$\text{N}_{\rm H}$ ($10^{22}$ atoms $cm^{-2}$)&$4.00^{+0.19}_{-0.10}$&\\
$\text{N}_{\rm H}^{\rm Si}$ ($10^{16}$ atoms $cm^{-2}$)&$2.45^{+0.17}_{-0.27}$&\\
\hline
& Diskbb&\\
$\text{kT}_{\rm in}$ (keV)&$0.49^{+0.06}_{-0.08}$&\\
 $\text{R}_{\rm in}$ (km)\footnote{Assuming a source distance of 4.5 kpcs and a mean inclination angle of $i = 60 \degr$ \citep{falanga}}& $18.08_{-11.92}^{+22.23}$&\\
\hline
& CompTT&\\
T0 (keV)&$0.82^{+0.03}_{-0.04}$&\\
kT (keV)&$16.27\footnote{Fixed at the best fitting value}$&\\
$\tau$&$2.30^{+0.09}_{-0.08}$&\\
Norm ($\times 10^{-2}$)&$3.82^{+0.18}_{-0.13}$&\\
\hline
& Emission line&\\
$\text{E}_{\rm l}$ (keV)&$6.72^{+0.10}_{-0.10}$&\\
$\sigma$ (keV)&$1.00^{+0.10}_{-0.10}$&\\
Norm ($\times 10^{-3}$) &$8.83^{+1.15}_{-1.01}$&\\
EW (keV)& 0.34&\\
\hline
&Cross-calibration&\\
DU1&$0.81^{+0.003}_{-0.003}$&\\
DU2&$0.81^{+0.003}_{-0.003}$&\\
DU3&$0.79^{+0.003}_{-0.003}$&\\
\hline
$\chi^{2}$/DOF& 1720/1484 \\
\hline
&Flux\footnote{Energy flux at 2-8, 2-4, and 4-8 keV energy ranges}  ($10^{-9}$ $ergs/cm^{2}/s$) &\\
 2-8 keV  &  $2.13^{+0.002}_{-0.001}$\\ 
2-4 keV & $0.82^{+0.001}_{-0.001}$\\ 
4-8 keV& $1.31^{+0.0006}_{-0.0005}$\\ 

\hline
$\text{F}_{\rm diskbb}^{\rm ph}$/$\text{F}_{\rm Total}^{\rm ph}$\footnote{Percentage of disk photon flux in the energy range 2–8 keV} (2-8)& 6.3\\
$\text{F}_{\rm diskbb}^{\rm ph}$/$\text{F}_{\rm Total}^{\rm ph}$\footnote{Percentage of disk photon flux in the energy range 2–4 keV} (2-4)&11.3\\
$\text{F}_{\rm diskbb}^{\rm ph}$/$\text{F}_{\rm Total}^{\rm ph}$\footnote{Percentage of disk photon flux in the energy range 4–8 keV} (4-8)&0.3\\
\hline
\end{tabular}
\label{table3}

\end{table}

\begin{table}
\centering
\caption{PD and PA of each spectral component obtained from the best-fit spectropolarimetric model {\tt tbnew*(diskbb*polconst+compTT*polconst+gauss)*const} and {\tt tbnew*(diskbb+compTT+gauss)*polconst*const} to the joint NICER and IXPE spectra of 4U~1728$-$34. The uncertainties are  at 90\% CL (see Section \ref{spec_pol_analysis}).}

\begin{tabular}{c c c}
\hline
Component & PD (\%)& PA (\degr)  \\ 
\hline
diskbb &$<40.3$&$\text{PA}_{\rm comptt}$\\
comptt &$<2.4$&$-44^{+15}_{-15}$\\
\hline
diskbb &$<87$&$\text{PA}_{\rm comptt}+90 \degr$\\
comptt &$1.9^{+1.1}_{-1.1}$&$-41^{+17}_{-17}$\\
\hline
dikbb+comptt &$1.9^{+1.0}_{-1.0}$&$-41^{+16}_{-16}$\\
\hline
\label{table4}
\end{tabular}

\end{table}

IXPE was launched on 2021 December 9 \citep{2022JATIS...8b6002W}. IXPE consists of three identical grazing incidence telescopes, each comprised of a NASA-provided Mirror Module Assembly (MMA) with an ASI-provided polarization-sensitive Detector Unit (DU) equipped with a gas-pixel detector. Each MMA consists of 24 concentrically nested nickel-cobalt-alloy mirror shells with a Wolter-1 configuration. IXPE is ideal for imaging and spectro-polarimetry in the 2–8 keV energy range \citep{2001Natur.411..662C,2021AJ....162..208S,2022JATIS...8b6002W}.

IXPE observed 4U~1728$-$34 from 2024 Aug 17, 08:59:42.184 UTC to Aug 19, 09:57:25.184 UTC with a total on-source exposure time of approximately 95.8 ks (see Table \ref{table1} and the light curve in Figure \ref{lc}). Spectral and polarimetric analysis was performed using {\tt HEASoft version 6.33}, with the IXPE Calibration Database (CALDB) version 20240125\footnote{\url{https://heasarc.gsfc.nasa.gov/docs/ixpe/caldb/}}. For extracting images and spectra, {\tt XSELECT} available as a part of the {\tt HEASoft 6.33} package was used extensively. Source photons were selected from circular regions with a radius of 60$\arcsec$ for I, Q, and U spectra for each DU centered at the brightest pixel located at RA of $262\fdg98$ and DEC of $-33\fdg83$. The background regions were defined as annuli with an inner radius of 132$\arcsec$ and an outer radius of 252$\arcsec$.  The weighted scheme  NEFF was adopted for the spectro-polarimetric analysis with improved data sensitivity\footnote{\url{https://heasarc.gsfc.nasa.gov/docs/ixpe/analysis/IXPE_quickstart.pdf}} \citep{2022SoftX..1901194B, 2022AJ....163..170D}.  The ancillary response files (ARFs)
and modulation response files (MRFs) were generated for each DU
using the {\tt ixpecalcarf} task, with the same extraction
radius used for the source region. The background rejection and subtraction schemes were implemented,
following the prescription by \cite{2023AJ....165..143D}. The unweighted model-independent polarimetric analysis was performed using the {\tt IXPEOBSSIM package version 31.0.1} \citep{2022SoftX..1901194B}. 

\subsection{NICER}
The Neutron Star Interior Composition Explorer (NICER) is an International Space Station (ISS) payload and is operating in the 0.2--12 keV energy range \citep{2016SPIE.9905E..1HG}. NICER's X-ray Timing Instrument (XTI) comprises a collection of 56 X-ray concentrator optics (XRC) and silicon drift detector (SDD) pairs \citep{2012SPIE.8443E..13G}.

NICER observed 4U~1728$-$34 from 2024 Aug 17, 16:21:26 UTC to 2024 Aug 19, 01:04:20 UTC. The observation details are summarized in Table \ref{table1}. The NICER/XTI observations were reduced using the NICER software NICERDAS distributed with {\tt HEASoft 6.33}, the Calibration Database (CALDB) 20240206\footnote{\url{https://heasarc.gsfc.nasa.gov/docs/heasarc/caldb/nicer/}}, and updated geomagnetic data. Cleaned event files were generated using the {\tt nicerl2} pipeline and by applying standard filtering criteria, limiting undershoot rates to $<$ 500 cts/s and overshoot rates to $<$ 30 cts/s. The {\tt nicerl3-spect} task was employed to generate source spectra and background spectra using the {\tt SCORPEON} background model, along with the detector responses. We note that the NICER observations were carried out during orbit day and orbit night. The orbit night observations comprised only 219 s on-source exposure time. Hence, we used orbit day data only for the spectro-polarimetric analysis.

\section{Results}
\label{sec4}
\subsection{Source Spectral State}
\label{spectral_analysis}

Figure~\ref{lc} shows the background-subtracted IXPE and NICER light curves of the source 4U~1728$-$34, which shows no strong variability in the X-ray flux during the observations. However, a total of 10 Type-I X-ray bursts were detected during the IXPE observations. The only burst detected during the NICER observation does not overlap with the IXPE observations. The source hardness in the IXPE data does not show any strong variations (not shown). Figure~\ref{hid} shows the NICER HID, where the hardness is defined as the ratio of the count rates between the 4.0-10.0 keV and 0.5-2.0 keV bands. The NICER HID indicates that the source was in a relatively hard state during the observations compared to past observations with NICER \citep{2023ApJ...958...55B}. To investigate the source spectral behavior during our observations further, we carried out detailed spectro-polarimetric investigations. 

It should be noted here that, we extracted the burst and non-burst data separately from the IXPE light curves and we use non-burst periods only for the spectro-polarimetric analysis.

\subsection{Model-independent polarimetric analysis}
\label{model_ind_analysis}

We performed polarimetric analysis of 4U~1728$-$34 using the {\tt ixpeobssim} package \citep{2022SoftX..1901194B} under the {\tt PCUBE} algorithm in the {\tt xpbin} tool \citep{2015APh....68...45K}. The unweighted analysis on the background-rejected (and not background-subtracted) data implemented in the {\tt ixpeobssim} package was applied. We employed the polarimetric analysis of the background-rejected non-burst data in the 2–8 keV, 2-4 keV, and 4-8 keV energy bands, and the results obtained are reported in Table \ref{table 2}. The minimum detectable polarization obtained for the non-burst interval at 99\% confidence is estimated as $\sim$ 2.4 \%.  Figure~\ref{polar_mod_ind} shows the
two-dimensional contour plots of the polarization degree (PD) and polarization angle (PA) at different energies obtained using the {\tt PCUBE} algorithm. We note here that the estimated uncertainties include background events, and the results of the {\tt PCUBE} analysis are not directly comparable to those resulting from the spectro-polarimetric analysis. The latter represents a more accurate analysis. 

Furthermore, the model-independent polarimetric analysis carried out for the burst-only interval summing up the data from the ten bursts detected during the IXPE observations, did not show any significant X-ray polarization. The corresponding minimum detectable polarization at 99\% confidence is estimated as $\sim$ 15 \%.

\subsection{Spectro-polarimetric analysis}
\label{spec_pol_analysis}
\begin{figure}
\centering
\includegraphics[width=0.50\textwidth]{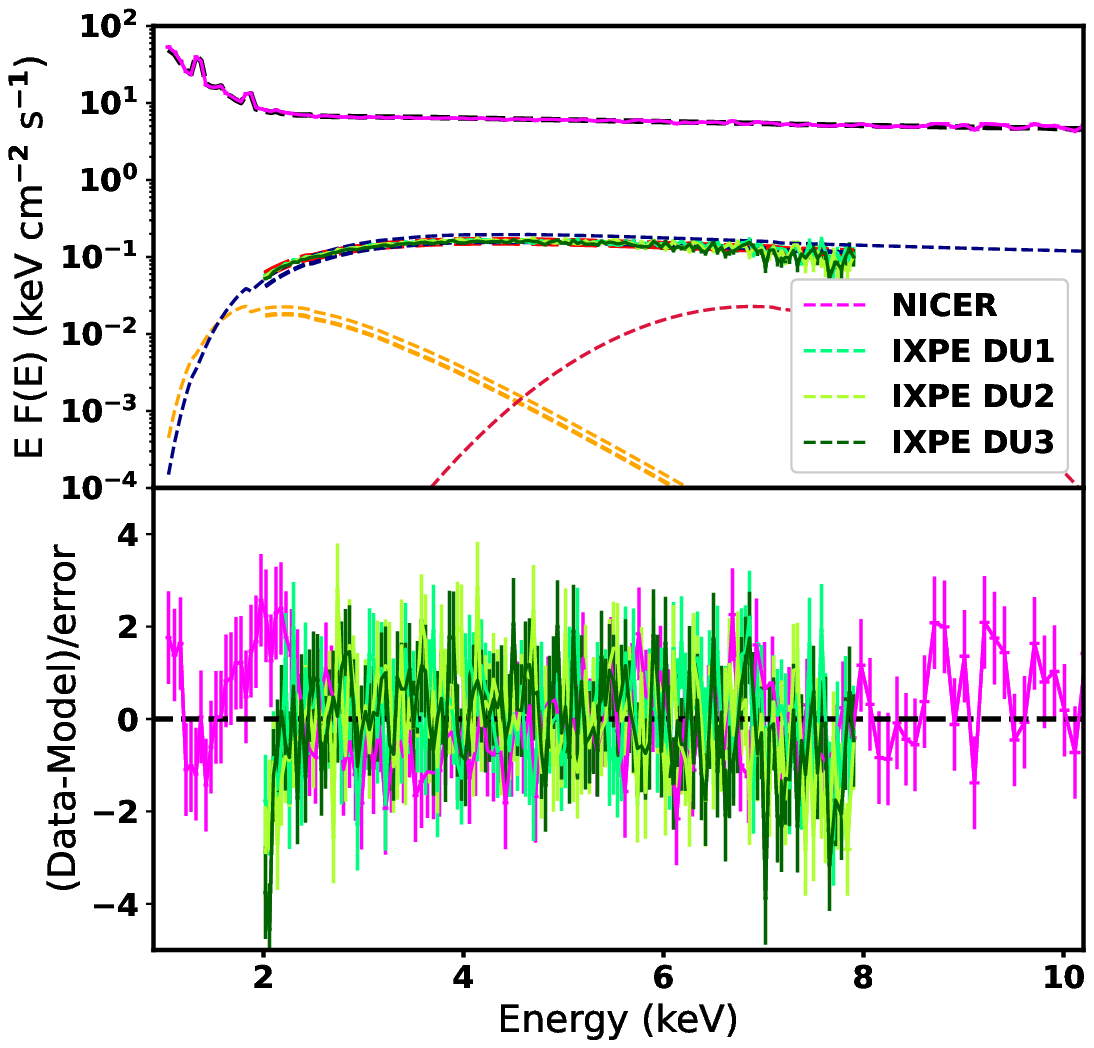}
\caption{Model fitted deconvolved joint spectra of 4U~1728$-$34 as observed by IXPE DU1 (spring green), IXPE DU2 (green yellow), IXPE DU3 (dark green), and NICER (magenta). The spectra are fitted with the {\tt tbnew*(diskbb+comptt)*polconst*const} model in the 1-15 keV energy band. The total model is shown with the dashed black (NICER) and red line (IXPE), and the individual additive components {\tt diskbb}, {\tt comptt}, and {\tt gauss} are shown with the dashed orange, dashed navy, and dashed crimson lines, respectively. The lower subpanel shows the residuals between the data and the best fit model.}
\label{spec_all}
\end{figure}

\begin{figure*}
\centering
\includegraphics[width=0.32\textwidth]{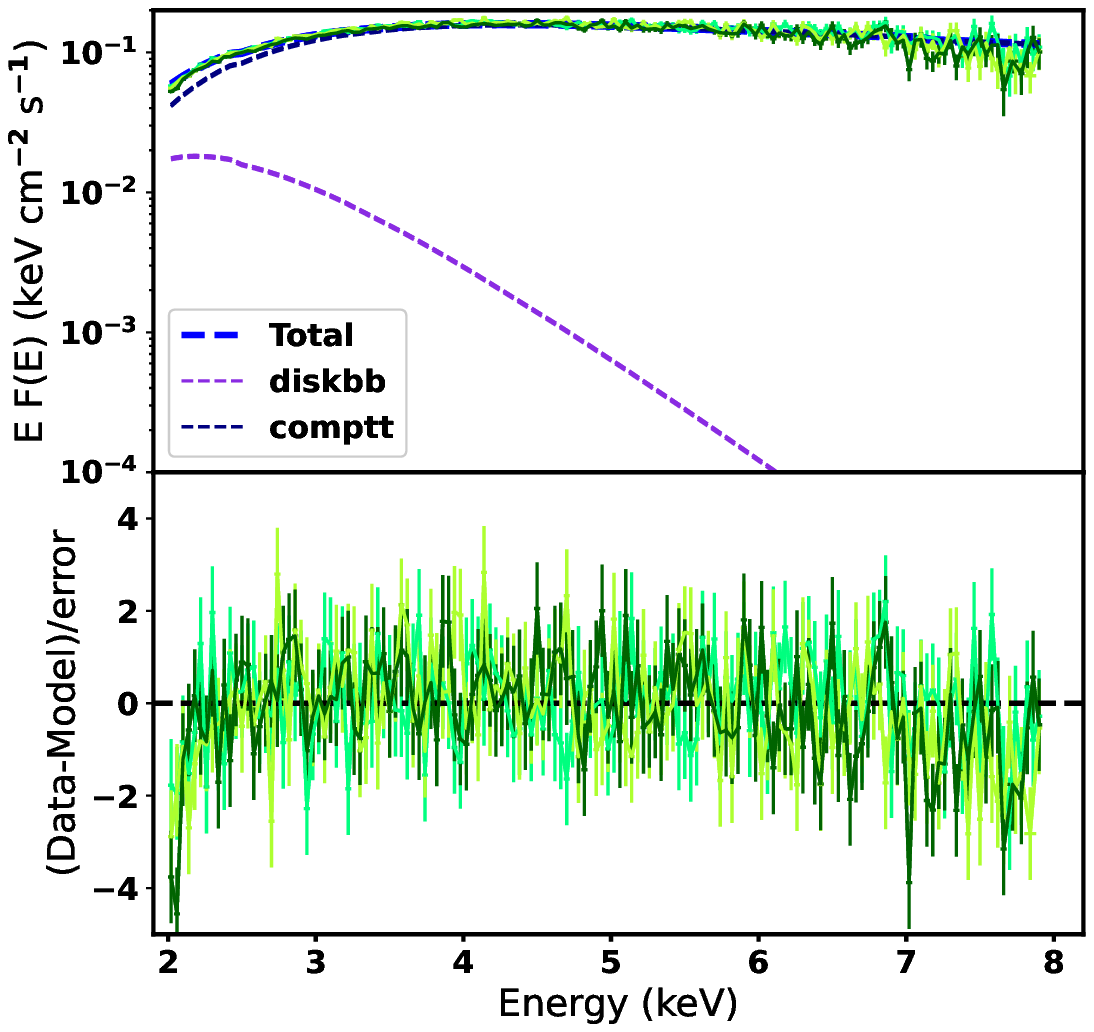}
\includegraphics[width=0.32\textwidth]{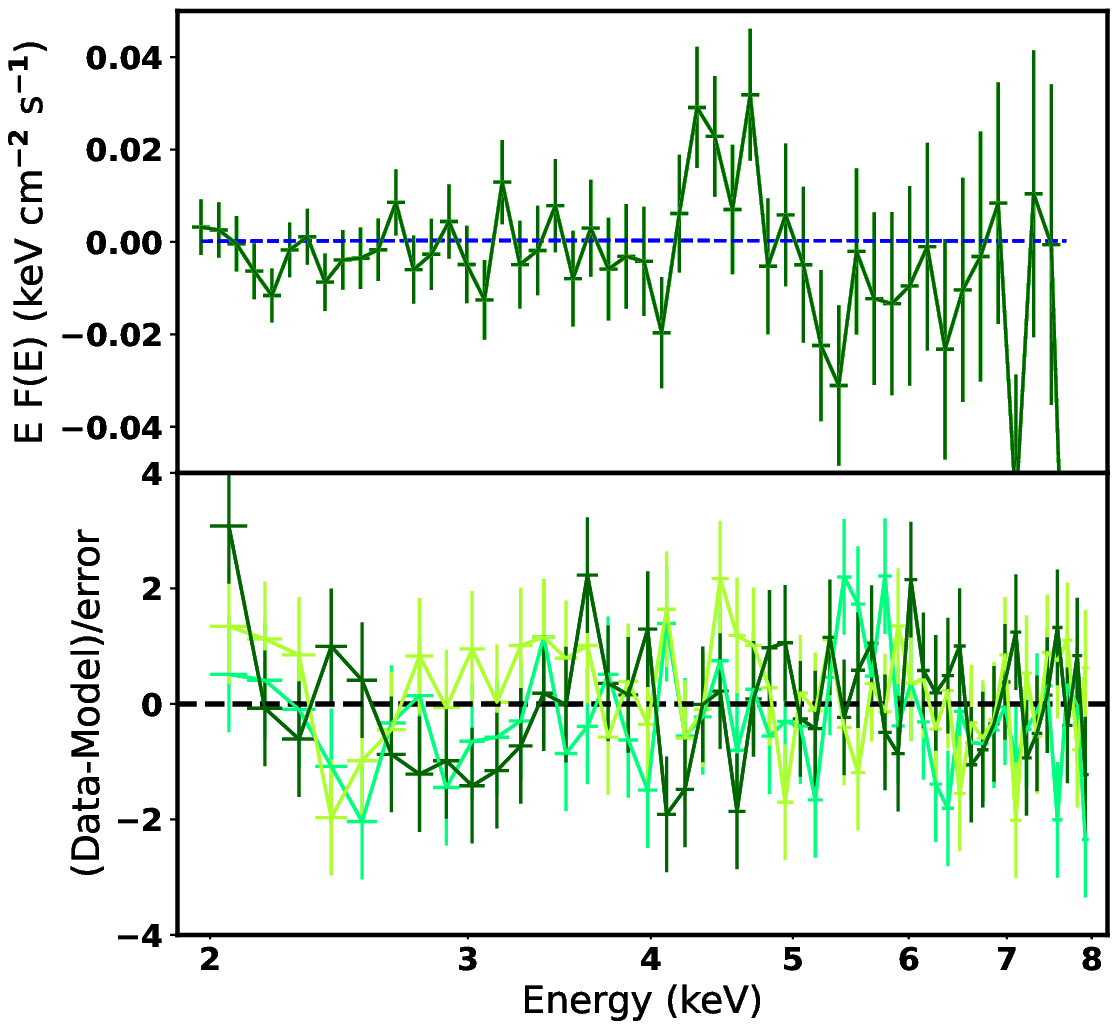}
\includegraphics[width=0.32\textwidth]{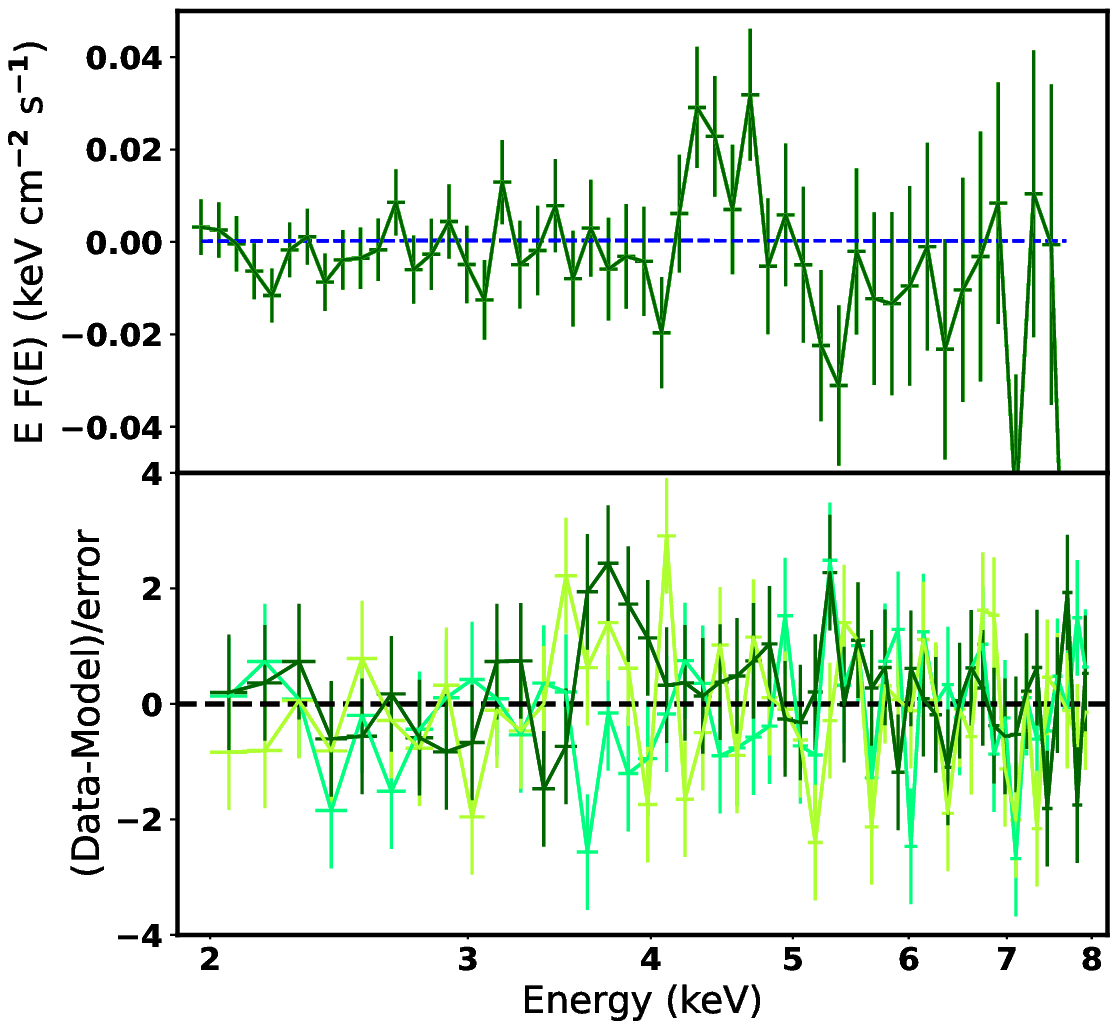}

\caption{Deconvolved spectra of 4U~1728$-$34 as observed by IXPE DU1 (spring green), IXPE DU2 (green yellow), IXPE DU3 (dark green) and the best fitting model {\tt tbnew*(diskbb+comptt)*polconst*const}  obtained from the joint NICER and IXPE analysis. The left, middle, and right panels show the Stokes I, Q, and U spectra, respectively. The total model is shown with the dashed blue line (left, middle and right). The individual additive components {\tt comptt} and {\tt diskbb} are shown with the dashed blue-violet and dashed navy lines, respectively. The lower subpanels show the residuals between the data and the best fit model. The data for the Q and U spectra are rebinned only for plotting and representation purposes.}
\label{spec_iqu}
\end{figure*}

 The spectral fitting and statistical analysis was carried out using the {\tt XSPEC v 12.14.0h} spectral fitting package distributed as a part of the {\tt HEASoft 6.33 package}.  Considering the entire IXPE observation and the NICER orbit day observation, we fitted the spectra with a model consisting of multicolour disk blackbody component ({\tt diskbb} in {\tt XSPEC}; \cite{1984PASJ...36..741M}) and a  Comptonization component ({\tt comptt} in {\tt XSPEC}; \cite{1994ApJ...434..570T}) in the 2.0–8.0 keV energy range (IXPE) and 1.0-15.0 keV energy range (NICER). The plasma temperature was kept frozen at its best-fit value, as otherwise fit parameters become completely unconstrained. The spectrum was modified by the presence of neutral hydrogen absorption in the interstellar medium, and this was taken care of by using the {\tt tbnew}\footnote{\url{https://pulsar.sternwarte.uni-erlangen.de/wilms/research/tbabs/}} model. In our spectral fits, we noticed a significant residual around $\sim$ 1.8~keV, which is likely the Si K edge, a known NICER instrumental systematic\footnote{\url{https://heasarc.gsfc.nasa.gov/docs/nicer/data_analysis/workshops/2024/joint2024.html}}. To account for this, we allowed for the Si abundance to vary relative to the \texttt{wilms} abundance, where we found a value of $2.45_{-0.17}^{+0.27}$ $\times 10^{16}$ atoms $cm^{-2}$, which improved the fit significantly. The iron (emission) feature detected in the NICER spectrum at $\sim$ 6.72 keV is well represented with an additional Gaussian ({\tt gauss}), with an equivalent width of $\sim$ 0.34 keV. A constant ({\tt const}) model was used to account for the uncertainties in cross-calibration uncertainties between NICER and the IXPE DUs and is reported in Table \ref{table3}. To study the polarization of the spectral components, we applied the {\tt polconst} model to the entire continuum model to compare with the results obtained from the model-independent analysis. The PD and the PA obtained are consistent with the upper limits of PD and PA obtained from the model-independent analysis (see Section \ref{model_ind_analysis}). Figure~\ref{spec_all} shows
the NICER (magenta) and IXPE (spring green, green
yellow, and dark green) spectra along with the best-
fitting models. Figure~\ref{spec_iqu} shows the spectro-polarimetric fits for the I, Q, and U Stokes parameters for all three IXPE DUs, and the corresponding best fitting values are reported in Table \ref{table3} and \ref{table4}. Finally, we applied a model that assumes different constant polarization for each spectral component. However, the PD of the {\tt diskbb} component remain unconstrained and the polarization estimations of the individual components could not be obtained due to limited data sensitivity. Thus, to reduce the statistical uncertainties further and have better constraints on the PA, we allowed to vary only the PA of the {\tt comptt} component, with the PA of {\tt diskbb} being linked by a relation $\text{PA}_{\rm diskbb}$=$\text{PA}_{\rm comptt}$ and $\text{PA}_{\rm diskbb}$=$\text{PA}_{\rm comptt}+90 \degr$. However, in both cases, the {\tt diskbb} polarization is essentially unconstrained by the data, with an upper limit of PD at the 90\% confidence level, it is $<40.3$\% for the former and $< 87.0$\% for the latter scenario(see Table~\ref{table4}). In the scenario where both the {\tt diskbb} and {\tt comptt} components are orthogonal in polarization angle, the constraints are slightly better on the {\tt comptt} component with a PD of $1.9^{+1.1}_{-1.1}$ at the 90\% confidence level. With the $\text{PA}_{\rm diskbb}$=$\text{PA}_{\rm comptt}$ setup, only upper limits could be obtained for the {\tt comptt} component ($<2.4$\%, at the 90\% confidence level) with the estimated PA of $-44^{+15}_{-15}$\%, at the 90\% confidence level. We report the result obtained from the spectro-polarimetric analysis in Table \ref{table3} and \ref{table4}. We also calculate the percentage of the total energy flux and {\tt diskbb} photon flux over the total in the 2-4, 4-8, and 2-8 keV energy bands (see Table \ref{table3}). It is thus clear that the data quality is insufficient for spectro-polarimetric decomposition. A more definitive statement would require a data set either with more counts or with a broader bandpass (as might be obtained with other polarization missions) so that the two spectral components can be disentangled properly.

\section{Radio Observations} 
\label{radio}

The Australia Telescope Compact Array (ATCA) observed 4U~1728$-$34 over three consecutive days in 2021 April in a hard X-ray spectral state. Full details of the light curves and short-time variability are presented in \citet{2024Natur.627..763R}. Here we discuss the polarization results of these data, which had not previously been presented.

The three ATCA observations occurred between 2021 April 3 and 2021 April 5, with the array in an extended 6\,km configuration (see Table \ref{table1}). Data were recorded simultaneously at central frequencies of 5.5\,GHz and 9\,GHz, with 2\,GHz of bandwidth at each frequency. We used PKS~B1934$-$638 for bandpass and flux calibration, and B1714$-$336 for phase calibration. Calibration and imaging were carried out following standard procedures within the Common Astronomy Software Applications ({\tt CASA}; \citealt{2022PASP..134k4501C}) package, version 5.1.3. Flux densities were measured by fitting a point source in the image plane.

To calibrate for polarization measurements, we used PKS B1934$-$638, which is unpolarized at our observing frequencies, to solve the antenna leakages (D-terms), and the {\tt CASA} task {\tt qufromgains} to solve the polarization angle of the phase calibrator. We do not detect linear or circular polarization at the position of 4U~1728$-$34 at either observing band. Taking the root mean square (rms) over the source position of the Stokes Q, U, and V images as 1 $\sigma$ errors, we place 3$\sigma$ upper-limits on the linear polarization (LP) of 2.0\% at 5.5\,GHz and 1.8\% at 9\,GHz, with circular polarization limits of 2.0\% and 1.8\%, respectively. Stacking the two bands together to provide our deepest limits gives a 3$\sigma$ upper limit on the LP of 1.5\% and 1.5\% on the circular polarization (measured at 7.25\,GHz). 

\subsection{Faraday rotation?}

Because 4U~1728-34 is located at a relatively large distance, deep within the Galactic Plane, and close to the Galactic Center, it may have strong Faraday rotation that makes estimation of its polarization challenging.  We can make a rough prediction of the amount of Faraday rotation for this position in the sky by using an estimate of the foreground electron column density \citep{2002astro.ph..7156C} along with an assumption about the Galactic magnetic field.  Taking a distance of 5 kpc gives a dispersion measure of 440 pc cm$^{-2}$.  Taking a Galactic magnetic field of 5 $\mu$G, which is likely to be higher than typical \citep{2024ApJ...970...95U}, the rotation measure will be about 1800 rad m$^{-2}$.  This is potentially a serious problem in our lower band, as this would give a rotation of close to $\pi$ radians between 4.6\,GHz and 6.4\,GHz.  Even in the higher band, between 8.1\,GHz and 9.9\,GHz, one would expect 0.8 radians of rotation (albeit with the likelihood that the weighted mean value of the Galactic magnetic field is smaller than 5 $\mu$G over this distance).

To test the possibility that Faraday rotation is responsible for the non-detection, we also separated each band into eight 256\,MHz sub-bands.  We do not find significant polarization in any of the sub-bands or grouped pairs of the sub-bands. Hence, to provide the deepest limits we report only the full bandwidth (5.5.\,GHz, 9\,GHz, and combined 5.5+9\,GHz bands) 3$\sigma$ upper-limits on polarization estimates as discussed above.

\section{Discussion}
\label{diss}
\begin{table*}
\centering
\scriptsize
\caption{The X-ray polarization properties of the NS LMXBs and hard state BH observed by IXPE during different spectral states, including polarization degree (PD), polarization angle (PA), the maximum detected polarization degree and the corresponding energy band, and the PA detected at the maximum PD (see Section \ref{diss}). }
\begin{tabular}{ c c c c c c}
\hline
Source (Type) & PD  (\%) (State) & PA (\degr) &Max PD (\%) & PA (\degr) & ref \\ 
&(2--8 keV)& &(Band in keV)& at Max PD\\
\hline
GS~1826$-$238 (Atoll)&$<$1.3 (High soft)&-&-&-&\cite{2023ApJ...943..129C}  \\
GX~9$+$9  (Atoll)& $1.4\pm0.3$ (soft)&-&$2.2\pm0.5$ (4--8)&-&\cite{2023AA...676A..20U, 2024Galax..12...43U}\\
4U~1820$-$303 (Atoll)&$<$1.3 (soft) &-&$10.3\pm2.4$ (7--8)&$-67\pm7$&\cite{2023ApJ...953L..22D}\\ 
4U~1624$-$49 (Atoll)&$3.1\pm0.7$ (soft) &$81\pm6$ & $6.4\pm2.0$ (6--8)&$83\pm9$&\cite{2024ApJ...963..133S}\\
Ser~X-1  (Atoll) &$<$2 (soft)&$-60\pm15$&$1.9\pm0.8$ (4--8)&$-64\pm12$&\cite{2024AA...690A.200U}\\
GX~3$+$1 (Atoll)& $<$1.3 (soft)&-&$<$1.8 (4-8)&-&\cite{2024arXiv241110353G}\\

Cyg~X-2  (Z) &$1.8\pm0.3$ (NB)&$140\pm4$&$2.79\pm1.68$ (4--8)&$134\pm18$&\cite{2023MNRAS.519.3681F}\\
XTE~J1701$-$462 (Z/Atoll) &$4.58\pm0.37$ (HB)&$-37.7\pm2.3$&Const&-&\cite{2023AA...674L..10C}\\
 &$0.65\pm 0.35$ (NB)&$-57\pm16$&Const&-&\cite{2023AA...674L..10C}\\
Sco~X-1 (Z) &$1.0\pm0.2$ (SA/FB)& $8\pm6$&Const&-&\cite{2024ApJ...960L..11L}\\
GX~5$-$1 (Z)&$3.7\pm0.4$ (HB) &$-9\pm3$ &$5.4\pm0.7$ (5-8)&$-14.0\pm3.7$&\cite{2024AA...684A.137F}\\
 &$1.8\pm0.4$ (NB/FB)&$-9\pm6$ &$2.6\pm0.7$ (5-8)&$-21.0\pm7.9$&\cite{2024AA...684A.137F}\\
Cir~X-1 (Z/Atoll) &$1.6\pm0.3$/$1.4\pm0.3$ (hard/soft) &$37\pm5$/$12\pm7$&Const&-&\cite{2024ApJ...961L...8R}\\
GX~13$+$1  (Z/Atoll)&$2.4\pm0.3$ (soft) &$28\pm3$  & Varies with time/energy&-&\cite{2024AA...688A.217B}\\
GX~340$+$0 (Z)&$4.02\pm 0.35$ (HB) &$37.6 \pm 2.5$& $5.8\pm1.2$ (6.3-8.0)&$48\pm6$&\cite{2024arXiv240519324B}\\
& $1.22\pm0.25$ (NB) & $38\pm6$ & $1.91\pm0.54$ (5-8) & $43\pm8$ & \cite{2024arXiv241100350B} \\
\\
&&{\bf Hard State BH} &&\\
\\
Cygnus~X-1 &$3.9\pm0.2$ (hard)&$-22\pm1$&-&-&\cite{2022Sci...378..650K}\\
&$3.8\pm0.3$ (hard)&$-25\pm2$&-&-&\cite{2024Galax..12...54D}\\

\hline
\end{tabular}

\label{table6}
\end{table*}

In this paper, we report the first X-ray and radio polarization study of the NS atoll source 4U~1728$-$34 using IXPE and ATCA. The source was in a relatively hard state during the observation exhibiting 10 Type-I bursts during the IXPE observation and 1 burst during the NICER observation, with no strong X-ray variability in the persistent emission \citep{2023ApJ...958...55B}. The MDP at 99\% confidence for the burst only interval, including the 10 bursts detected by IXPE, is estimated to be $\sim$ 15\%. We also note that during our ATCA observations of 4U~1728$-$34, 10 type-I bursts were detected, resulting in bright radio flares associated with increased mass flow into the jet (see \cite{2024Natur.627..763R} for details). As the source 4U~1728$-$34 is a frequent burster, future IXPE observations with longer exposures or stacking all the archival data of any single source might help put better constraints on the burst polarization with greater burst statistics. 

 The non-burst X-ray spectrum of 4U~1728$-$34 is well described by a {\tt diskbb} component representing the thermal/softer emission ($kT_{\rm in}$ $\sim$ 0.49 keV), along with a non-thermal {\tt comptt} component representing the Comptonization of the thermal seed photons (T0 $\sim$ 0.82 keV) from a NS surface by hot electrons ($kT_{e} \sim$ 16) in an inner accretion flow (or outer BL) of relatively low optical depth ($\tau \sim$ 2.3), consistent with the previously reported spectrum of the source \citep{2006A&A...458...21F}. The comptonizing region with a relatively higher plasma temperature ($kT_{e} \sim$ 16) and a relatively low optical depth ($\tau \sim$ 2.3) may indicate a relatively hard state observation of the source (see hard/island state ``state C" observation of 4U~1820$-$30 \citep{2007ApJ...654..494T}). Assuming a source distance of 4.5 kpcs, the ratio of the colour to effective temperature $f_{\rm col} = 1.8$ and a mean inclination angle of $i = 60 \degr$ \citep{2006A&A...458...21F}, we find an inner disk radius of $R_{in} = 18.08_{-11.92}^{+22.23}$ km, which may translate to a inner disk radius of a truncated accretion disk in a hard state. Furthermore, the relatively higher energy flux estimated in the 4-8 keV energy band than that of the 2-4 keV energy band, where {\tt diskbb} component is expected to be dominating, along with the estimated low fraction of the {\tt diskbb} photon flux to the total flux may emphasize on the source being in a relatively hard state during our observations. We note here that the values reported in this work are not precise, but approximate estimations of the spectral parameters because of data limitations and a more accurate determination of the source spectral states requires a broadband spectral analysis using instruments with high-energy coverage. Additionally, to further disentangle the geometry of 4U~1728$-$34, an energy-dependent study is necessary. Due to the limited sensitivity and limited energy coverage of our observations, the energy dependence of polarization properties could not be examined further. The hard X-ray polarimetry missions such as XPoSat\footnote{\url{https://www.isro.gov.in/XPoSat_X-Ray_Polarimetry_Mission.html}}, balloon-borne missions such as XL-Calibur \citep{2021APh...12602529A}, and next-generation polarimetry detector such as the LargE Area burst Polarimeter (LEAP) \citep{2021AAS...23713502W}, and POLAR–2 \citep{2023arXiv230900518P} may address the current limitations, enabling precise energy dependent studies as well as broadband spectro-polarimetry.

 The X-ray spectro-polarimetric analysis of 4U~1728$-$34 shows an overall polarization of ${\rm PD} = 1.9 \pm 1.0\%$ with a polarization angle of ${\rm PA} = -41 \pm16 \degr$, consistent with the upper limits obtained from the model-independent polarimetric studies. IXPE has so far observed fourteen NS LMXBs, including 4U~1728$-$34, being the first NS atoll observed in a relatively hard state. Table \ref{table6} shows the previously reported PD and PA of the NS LMXBs observed by IXPE. Typically, the X-ray spectrum of the weakly magnetized NS (WMNS) is thought to be comprised of two main components: the softer component is interpreted as the emission from the disk and the harder component from the BL or the SL \citep{2024arXiv240916023B}.   In the case of 4U~1728$-$34, we obtain the PD associated with the  Comptonization region ${\rm PD} = 1.9 \pm 1.1\%$, when the {\tt diskbb} component is considered perpendicular to the {\tt comptt} component; and an upper limit of PD  $<$ 2.4\%, when {\tt diskbb} component and {\tt comptt} component are aligned. However, in both cases,  the soft disk PD remains unconstrained.

 The previous studies of atolls show hints of disk polarization being consistent with the theoretical predictions for a semiinfinite plane-parallel atmosphere \citep{1960ratr.book.....C}. The polarization from the harder emission component in the case of NS atoll sources is reported to be associated with the Comptonization of the seed photons in a BL or a SL, and/or the reflection off the accretion disk  (see Table~\ref{table6} and the references therein). Although, from the current analysis, we cannot rule out one scenario in favor of the others, the feature detected in the NICER spectrum of 4U~1728$-$34 at $\sim$ 6.7 keV fitted with a Gaussian model of equivalent width $\sim$ 0.34 keV, indicates the presence of a broad iron emission line, thereby confirming the presence of reflected emission (see \cite{2017MNRAS.468.2256W,2019MNRAS.484.3004W} and the references therein).

  Compton scattering may result in high polarization in the IXPE bandpass if the cold photons are upscattered in the hot Comptonizing regions. However, in the case of WMNS, the SL is expected to be at a low temperature of 2–3 keV, and the photons are expected to be observed in the same energy range, regardless of the number of scatterings. Thus, in the case of SL, an overall low polarization is expected regardless of the optical depth \citep{2024arXiv240916023B}.  The estimated PD of $1.9^{+1.1}_{-1.1}$\% from the Comptonizing region of 4U~1728$-$34, when the {\tt diskbb} and {\tt comptt} components are considered to be perpendicular to one another, may imply an SL-like geometry of the source, consistent with the predicted simulation results considering the current SL model in  \cite{2024arXiv240916023B}. A strong PA dependency on energy is also observed if the harder component comes from the SL rather than the BL, which has similar polarization to the disk \citep{2024arXiv240916023B}.  In the case of 4U~1728$-$34, we do not see any such PA dependency on energy.  However, given the sensitivity of the data presented in this work, it is not possible to rule out the SL model, based on the results obtained from the energy-resolved polarization study. However, the relatively higher temperature obtained for the Comptonization region ($kT_{e} \sim$ 16) during our hard state observation of  4U~1728$-$34 than the expected range of temperature of the SL, suggests that an SL scenario may not be plausible in this case.

  On the other hand, assuming a source inclination of 60\degr, a BL coplanar with the accretion disk can also explain the observed  PD of $< 2.4$\% with the PA ($\sim$ $-44^{+15}_{-15}$) of the harder emission component representing the BL aligned with the softer disk emission. Considering the results reported in Figure 5 of \cite{1985A&A...143..374S}, the measured PD of $<$ 2.4\% is compatible with a slab of a Thomson optical depth of $\tau = 2.3$ (See table \ref{table4}).

Typically, the source geometry is thought to vary as the source transitions between different X-ray spectral states \citep{2007A&ARv..15....1D}. Even though our results are consistent with the previously reported polarization of NS atoll sources, this study, being the first estimate of polarization of a relatively hard state atoll, can not make a direct comparison with the previous estimations.  At present, 4U~1728$-$34 is the only NS atoll source observed with IXPE in the hard state (see Table \ref{table6}). Typically, the hard state geometry is expected to be different from the previously investigated geometry of the soft state atoll sources reported by IXPE. In the hard state, at a low accretion rate, the disk is thought to be truncated, while the hot inner ﬂow joins smoothly onto the optically thin BL \citep{2001AIPC..587...54M,2004ApJ...613..506M}. As the source moves from the hard to the soft state, the disk approaches the NS with an increasing accretion rate, and the hot inner ﬂow collapses into a thin disk. Consequently, the BL becomes optically thick to electron scattering. Thus, the BL spectrum changes dramatically from hot, optically thin Comptonization to much lower temperature, optically thick Comptonization, as the source transitions from the hard island state to the soft banana branch in the case of NS atolls. Hence, the polarization contribution from the hard state optically thin BL is expected to be relatively higher than that of the soft state optically thick BL, as the increased numbers of scattering may lead to reduced polarization in a region of higher optical depth.

Considering the fact that the island state of the atoll has its counterpart with the HB of Z sources, we compare our results with the previously reported polarization study of the Z sources in the HB. Our estimated upper limit of the PD of the Comptonization region is relatively smaller than the previously reported HB state observation of NS Z sources \citep{2023AA...674L..10C,2024AA...684A.137F,2024arXiv240519324B}. 

 Arguments are often made that the hard states of   NS LMXBs are similar to the hard states of  BH sources on the basis of both spectroscopy \citep{2002MNRAS.337.1373G} and timing \citep{1994A&A...283..469V}.  At present, Cyg X-1, being the only hard state BH X-ray binary observed by IXPE, is the only source that can be a possible candidate for such comparison \citep{2022Sci...378..650K}. The observed alignment of the X-ray polarization in Cyg X-1 with the radio jet reveals the presence of the spatially extended hot X-ray emitting plasma in a plane perpendicular to the jet axis. As the similarities between neutron stars and black holes may be because of similar emission mechanisms \citep{Medvedev}, or may be due to the effects of a boundary layer \citep{Popham}, it is of great interest to compare the two classes of systems in order to understand the neutron star emission geometry.

ATCA radio data presented in this work provide a 3$\sigma$ upper limit on the linear polarization of 2.0\% at 5.5 GHz and 1.8\% at 9 GHz associated with jets detected during the hard X-ray state of 4U~1728$-$34. Stacking the two bands provides the deepest limit at a 3$\sigma$ upper limit on LP of 1.5\%. Theoretically, the radio emission from the hard state jets can be linearly polarized up to $\sim$10\%  when the magnetic fields are highly ordered \citep{1979rpa..book.....R,2011hea..book.....L}.  However, in practice these levels of LP are rarely observed in LMXBs where tangled magnetic fields \citep{2011arXiv1104.0837R}, the emission arising from multiple components \citep{2004MNRAS.354.1239S}, or spatially dependent Faraday rotation \citep{2013MNRAS.432..931B} reduce the fractional polarization signature. As such, our constraints are well within the expected levels for the jet emission from an LMXB, which are typically observed at levels of a few percent at most for the hard state jet (e.g., \citep{2015MNRAS.450.1745R}). 

The radio LP in both NS and BH LMXBs has been observed to vary as a result of changes in the jet emission (increased mass accretion rate, ejection events, changes in the structure), which can in turn better order the magnetic field, increasing the observed fractional LP \citep{2013MNRAS.432..931B}. The estimated LP of the radio emission from 4U~1728$-$34 reported in this work provides a constraining upper limit of the LP of the radio emission from 4U~1728$-$34 during a hard spectral state. Further investigations of NS LMXBs are crucial to investigate the presence of variable LP in NS LMXBs and its correlation with the accretion rate or spectral state of the systems. As a follow-up to this work, we have a planned more sensitive radio observation of 4U~1728$-$34 with the Very Large Array in the near future. More sensitive observations in the radio band may also provide a precise measurement or even tighter constraint on the LP, possibly revealing the position angle of the radio jet from 4U~1728$-$34 \citep[although the agreement between the jet orientation and polarization angle may not always hold, see discussions in e.g.,][]{2015MNRAS.450.1745R}.

\section{Summary}
In this work, we report the first X-ray and radio polarization study of 4U~1728$-$34, the first hard state NS atoll source observed with IXPE. The X-ray spectro-polarimetric study using IXPE and NICER shows a source polarization of  ${\rm PD} = 1.9 \pm 1.0\%$  with a polarization angle of ${\rm PA} = -41 \pm16 \degr$. The detected polarization of the Comptonizing region represents a BL-like geometry, typical of atoll type NS and WMNS. We also report the radio polarization of 4U~1728$-$34 using ATCA, with a 3$\sigma$ upper limit of 2.0\% at 5.5 GHz and 1.8\% at 9 GHz, and the deepest upper limits of $<1.5$ \% on linear and circular polarization at 7.25 GHz. Future monitoring in both the radio band and X-ray band
with longer exposure would enable polarimetric analysis
throughout different accretion states of the source allowing us to disentangle the geometry of different components. This study also demands more observations of so far limited hard state sources using IXPE, including both BH and NS X-ray binaries to better test their predicted geometry.

\section{Acknowledgments}
\begin{acknowledgments}
This research used data provided by the Imaging X-ray Polarimetry Explorer (IXPE), NICER (Neutron star Interior Composition Explorer), and ATCA (Australia Telescope Compact Array) and distributed with additional software tools by the High-Energy Astrophysics Science Archive Research Center (HEASARC), at NASA Goddard Space Flight Center (GSFC). U.K. and T.J.M.  acknowledges support by NASA grant 80NSSC24K1747. The ATCA is part of the Australia Telescope National Facility (\url{https://ror.org/05qajvd42}) which is funded by the Australian Government for operation as a National Facility managed by CSIRO. We acknowledge the Gomeroi people as the Traditional Owners of the ATCA observatory site. T.D.R. is an INAF IAF research fellow. M.N. is a Fonds de Recherche du Quebec – Nature et Technologies (FRQNT) postdoctoral fellow.

\end{acknowledgments}

\vspace{5mm}
\facilities{IXPE, NICER, ATCA}

\software{ixpeobssim \citep{2022SoftX..1901194B}, xspec \citep{1996ASPC..101...17A}, HEASoft \citep{1995ASPC...77..367B}
          }

\bibliography{rev}{}
\bibliographystyle{aasjournal}

\end{document}